\documentstyle[preprint,aps,epsfig]{revtex}
\tightenlines
\begin{document}
\draft
\preprint{\begin{minipage}[b]{1.5in}
          UK/TP 98-13\\
          BNL-HET-98/36
          \end{minipage}}
\vspace{0.2in}

\title{The Leading Power Corrections to the Structure Functions}
\author{Xiaofeng Guo$^1$  and Jianwei Qiu$^{2,3}$}
\address{$^1$Department of Physics and Astronomy, 
             University of Kentucky,\\
             Lexington, Kentucky 40506, USA\\
         $^2$Department of Physics and Astronomy, 
             Iowa State University \\
             Ames, Iowa 50011, USA \\
         $^3$Physics Department, 
             Brookhaven National Laboratory\\
             Upton, New York 11973-5000, USA }
\date{October 29, 1998}
\maketitle

\begin{abstract}
We show that when Bjorken $x_B$ is large, the leading order twist-four 
contributions to the structure functions can be expressed in terms of 
the derivatives of the normal twist-two parton distributions.  Our
analytical results are not only consistent 
with the parameterized power corrections extracted from the data, 
but also predict the flavor dependence of the power
corrections.  We also discuss the impact of our analytical results on
the extraction and the flavor separation of the parton
distributions at large $x$. 
\end{abstract}
\vspace{0.2in}

\section{Introduction}

Perturbative Quantum Chromodynamics (QCD) has been very successful in
interpreting and predicting scattering cross sections at large
momentum transfer.  For hadronic collisions, a number of sets of
reliable parton distributions are available for perturbative
calculations of potential experimental signals \cite{PDF}.  However, 
recent data on inclusive high $P_T$ jets from Fermilab \cite{HTjet}
and high $Q^2$ events at HERA \cite{HQdis} remind us that there are
still large uncertainties in the parton distributions at large $x$. 
The precise knowledge of parton distributions at large $x$ is very
important in search for signals of new physics at very high
energies, which might be reached only by the partons with large
momentum fraction $x$ at the colliders.  

According to the QCD perturbation theory, parton distributions at large
$x$ and high $Q^2$ can be derived from the parton distributions at
large $x$ and lower $Q^2$ by solving the DGLAP evolution equations.  
However, data from most physical observables are only sensitive to the
parton distributions in the intermediate or small $x$ region, because
of the phase space integration and the fact that the parton
distributions decrease very fast as $x$ increases.  It is believed
that structure functions measured in deeply inelastic scattering (DIS)
at large Bjorken $x_B$ and low $Q^2$ might be the best source to get
the direct information on the parton distributions at
large $x$.  But, at large $x_B$ and low $Q^2$, the structure functions
receive significant power corrections \cite{YB1}, and these power
corrections confuse the determination of the parton distributions.  

In general, there are two types of power corrections, known as the
kinematic and the dynamical power corrections.  The kinematic power
corrections is due to the fact that the nucleon has a finite mass, $m_N$.
Such power corrections have been systematically discussed in 
Ref.~\cite{Georgi}. On the other hand, the dynamical power corrections
are the results of the interactions between quarks and gluons
\cite{Georgi,Jaffe,DW}.  The dynamical power corrections are in general
proportional to the multiparton correlation functions or the matrix
elements of high twist operators, which are different from the normal
parton distributions.  Because of the large number of unknown
multiparton correlation functions, the actual size of the dynamical
power corrections to the structure functions have not been well
understood theoretically.  In this paper, we derive analytically the
leading dynamical power corrections to the structure functions, and
provide the numerical estimates of the power corrections.  Since we
will not discuss the kinematic power corrections in this paper,
without any confusion, the words ``power corrections'' in the rest of
this paper indicate the ``dynamical power corrections''.  

In a spin-averaged lepton-hadron deeply inelastic scattering,
$l(k)+h(p)\rightarrow l(k')+X$, the hadronic tensor $
W^{\mu\nu}(x_B,Q^2)$ can be decomposed as  \cite{CTEQhk}
\begin{eqnarray}
W^{\mu\nu}(x_B,Q^2) &= &
-\left( g^{\mu\nu} -\frac{q^\mu q^\nu}{q^2} \right)\, F_1(x_B,Q^2)
\nonumber \\
&+& \frac{1}{p\cdot q}\,
\left(p^\mu - \frac{p\cdot q}{q^2}\, q^\mu\right)
\left(p^\nu - \frac{p\cdot q}{q^2}\, q^\nu\right)\, F_2(x_B,Q^2)\ ,
\label{w}
\end{eqnarray}
where $q=k-k'$ is the four-momentum of the virtual photon, $q^2 =
-Q^2$, and $F_i(x_B,Q^2)$ with $i=1,2$ are the structure
functions.  At the large enough $Q^2$, the 
structure functions can be factorized into a convolution of the
short-distance coefficient functions and nonperturbative matrix
elements of multiparton fields \cite{CTEQhk}, 
\begin{equation}
F_i(x_B,Q^2)=\sum_{j,m} C^{(j,m)}_i(x_B,x_1,x_2,..,Q^2/\mu^2)
\otimes T^{(j,m)}(x_1,x_2,..,\mu^2)\left(\frac{1}{Q^2}\right)^m,
\label{Fxq}
\end{equation}
where $x_i$ with $i=1,...$ are parton momentum fractions, $\mu^2$ is
the factorization scale and is often chosen to be $Q^2$ in DIS,
$C^{(j,m)}$ are the perturbatively calculable coefficient functions, 
and $T^{(j,m)}$ are the corresponding nonperturbative matrix elements.
In Eq.~(\ref{Fxq}), $\sum_m$ runs from 0 to infinity, and $\sum_j$
sums over all possible types of nonperturbative matrix elements at a
given power of $m$.  The term with $m=0$ is often known as the twist-2 
or leading power contributions, and the terms with $m\neq 0$ are the
high twist contributions or power corrections.  The factorization
presented in Eq.~(\ref{Fxq}) is useful only when the terms of high
power corrections can be neglected.  When the physical scale, $Q^2$,
is large enough, the leading power contributions are often sufficient,
and the structure functions can be expressed in terms of a few  
nonperturbative matrix elements, $T^{(j,0)}(x,Q^2)$.  These matrix
elements $T^{(j,0)}(x,Q^2)$ with $j=q,\bar{q}$ and $g$ are known as
the normal quark, antiquark, and gluon distributions, and have been  
well-determined for a wide range of $x$ and $Q^2$ \cite{PDF}.  

Working in the moment space of the structure functions, $F_i(n,Q^2) =
\int_0^1 dx\, x^n\, F_i(x,Q^2)$, De R\'ujula, et al. \cite{Georgi}
demonstrated that when the moment $n$ is large enough, 
\begin{equation}
F_2(n,Q^2) \sim \sum_m \left[\frac{n\, \mu_0^2}{Q^2}\right]^m\,
T^m(n,Q^2)\ ,
\label{moment}
\end{equation}
with $\mu_0^2$ a non-perturbative mass scale, and concluded that there
exist a region of $n\sim Q^2/\mu_0^2$ where the power corrections are
as important as the leading contributions. This conclusion  
is equivalent to say that in the $x_B$-space, when $x_B\rightarrow 1$,
high order terms in the expansion in Eq.~(\ref{Fxq}) become important
even if $Q^2$ is large.
Physically, when $x_B\rightarrow 1$, the invariant mass of the
photon-hadron system in DIS approaches to the resonance region, and
the perturbative expansion in Eq.~(\ref{Fxq}) is then invalid.  

In order to avoid the resonances, we have to require the invariant
mass of the photon-hadron system \cite{Bodek,Ji}
\begin{equation}
W^2\equiv (p+q)^2 = m_N^2 + \frac{1-x_B}{x_B}\, Q^2 
> 2\ \mbox{GeV}^2\ .
\label{W2}
\end{equation} 
On the other hand, in order to extract the information on parton
distributions at large $x_B$ and low $Q^2$, we have to work in a
region where $W^2$ is small and not too far away from the resonances,
and will have to deal with the power corrections.  The purpose of this
paper is to get a good estimate of the size of the leading power  
corrections and to discuss its impact on the extraction of the
reliable twist-2 parton distributions at large $x$. 

By fitting DIS data in large $x_B$ region, it has been found
\cite{YB1,YB2} that the $1/Q^2$ dynamical power corrections to the
structure functions can be parameterized in terms of the normal parton
distributions,
\begin{equation}
F_2(x_B,Q^2) \approx \left(1+\frac{h(x_B)}{Q^2} \right)
F_2(x_B,Q^2)_{{\rm LT}}\, ,
\label{f2}
\end{equation}
where $F_2(x_B,Q^2)_{{\rm LT}}$ includes all leading twist
(or leading power) contributions as well as the target mass
corrections to the full structure function $F_2(x_B,Q^2)$.  The $h(x)$
in Eq.~(\ref{f2}) is a phenomenological fitting function, and is 
parameterized as \cite{YB2} 
\begin{equation}
h(x_B) = a \left(\frac{x_B^b}{1-x_B} - c\right) \, ,
\label{hx}
\end{equation}
where $a,b$ and $c$ are constant fitting parameters.
Because of the $1/(1-x_B)$ dependence in Eq.~(\ref{hx}), it is clear
that the dynamical power corrections in Eq.~(\ref{f2}) become more
important when $x_B$ increases.  It has been argued that the 
simple parameterization in Eq.~(\ref{hx}) can provide a very good fit
to the existing DIS data on proton as well as deuteron targets (with
slightly different fitting parameters), and demonstrated that  
the power corrections are important for extracting the correct parton
distributions, such as $d/u$ ratio, in large $x$ region
\cite{YB1,YB2}. 

On the other hand, a complete expression of the $1/Q^2$ power
corrections to the structure functions at the leading order of
$\alpha_s$ were derived more than fifteen years ago \cite{EFP,Qiu}.
As expected from Eq.~(\ref{Fxq}), the $1/Q^2$ corrections are
proportional to the matrix elements of four-parton operators, which
are very different from the operators defining the normal parton
distributions.  The success of the phenomenological parameterization
in Eq.~(\ref{f2}) raises a question: under what conditions, one can
derive the approximate expression in Eq.~(\ref{f2}), or a similar
expression, from the complete $1/Q^2$ power corrections at the leading
order of $\alpha_s$, which were derived in Ref.~\cite{EFP,Qiu}.   

In this paper, we show that when $x_B$ is large, the $1/Q^2$
power corrections to the structure functions at the leading order of
$\alpha_s$ can be approximated as  
\begin{equation}
F_2(x_B,Q^2) \approx \sum_{q,\bar{q}}\, e_q^2\, x_B 
\left\{1
     +\frac{1}{Q^2}\left[
      \langle D_T^2\rangle
      \left(-x_B \frac{d}{dx_B} \right)
      + 4\,\langle m_T^2\rangle\,
      \right]
\right\}\, q(x_B,Q^2)\ ,
\label{f24}
\end{equation}
where $m_N^2=p^2$ is the nucleon mass, $\langle D_T^2 \rangle$ and 
$\langle m_T^2 \rangle$ are the averaged values of $D_T^2$ and 
$m_T^2$, which will be defined in the next section.  Numerically, we  
demonstrate that our analytical expressions for the leading power
corrections in Eq.~(\ref{f24}) have the same $x_B$-dependence as the
phenomenological parameterization given in Eqs.~(\ref{f2}) and
(\ref{hx}), and therefore, we expect Eq.~(\ref{f24}) to fit the data
as well.  Instead of three unknown parameters, $a,b$ and $c$ in
Eq.~(\ref{hx}), our analytical result in Eq.~(\ref{f24}) have only two
parameters, and both of them have clear physical interpretations.
Furthermore, because of the scaling violation of the quark
distributions, the derivative in Eq.~(\ref{f24}) gives the natural
explanation for the flavor dependence (or target dependence) of the
fitting parameters in Eq.~(\ref{hx}) \cite{YB2}.  In addition, the
derivative provides some extra logarithmic $Q^2$-dependence for the
power corrections. 

Our result in Eq.~(\ref{f24}) demonstrates that although the power
corrections to the structure functions in DIS are in general very
complicated, the leading contributions in $\alpha_s$ at large $x_B$
are approximately given by the derivatives of the normal quark
distributions.  Such a simple analytical expression will be very
useful for the QCD global analysis of extracting reliable leading
twist parton distributions. 

The rest of our paper is organized as follows.  For the completeness,
we briefly derive the leading twist-4 contributions to the structure functions
in Sec.~\ref{sec2}.  In Sec.~\ref{sec3}, we identify the necessary
approximations, and derive the main result of this paper,
Eq.~(\ref{f24}), from the complete leading power corrections obtained
in Sec.~\ref{sec2}.  Finally, in Sec.~\ref{sec4}, we numerically
compare our analytical expression in Eq.~(\ref{f24}) with the
phenomenological parameterization in Eqs.~(\ref{f2}) and (\ref{hx}),
and estimate the values of the parameters $\langle D_T^2\rangle$ and
$\langle m_T^2\rangle$ in Eq.~(\ref{f24}).  We also discuss the
possible impact of our analytical results of the leading power corrections
to the extraction and the flavor separation of parton distributions at
large $x$. 

\section{Power corrections in DIS at leading order in $\alpha_s$}
\label{sec2}

The complete $1/Q^2$ power corrections to the structure functions at
the leading order of $\alpha_s$ can be separated into two groups: one
is proportional to the two-quark-two-gluon matrix elements (or
correlation functions), and the other is proportional to the
four-quark matrix elements \cite{EFP,Qiu}.  Because of Fermi
statistics and the fact that sea quark distributions are smaller than
the gluon distributions, contributions from the four-quark
subprocesses in large $x_B$ region should be smaller than that from
the quark-gluon subprocesses.  
In this section, we neglect the four-quark contributions, and drive
the leading order $1/Q^2$ 
contributions to the DIS structure functions from the quark-gluon
subprocess, from which the main result of this paper, Eq.~(\ref{f24}),
will be deduced in the next section.

For the definiteness of our derivations, we choose the following 
photon-hadron frame
\begin{eqnarray}
q^\mu &=& - x_B p\cdot n\, \bar{n}^\mu 
          + \frac{Q^2}{x_B\,p\cdot n}\, n^\mu \ ,
\nonumber \\
p^\mu &=& p\cdot n\, \bar{n}^\mu \ ,
\label{frame}
\end{eqnarray}
where we have neglected the hadron mass for the dynamical power
corrections.  In Eq.~(\ref{frame}), the four-vector, $\bar{n}^\mu$ and
$n^\mu$ are defined as 
\begin{eqnarray}
\bar{n}^\mu &\equiv & (\bar{n}^+,\bar{n}^-,\bar{n}_T) 
               = (1,0,0_T)\ ,
\nonumber \\
n^\mu &=& (0,1,0_T)\ , 
\label{nnbar}
\\
\bar{n}^2 &=& 0, \quad n^2 = 0, \quad \mbox{and} \quad
\bar{n}\cdot n = 1 \ .
\nonumber 
\end{eqnarray}
In terms of the $\bar{n}^\mu$ and $n^\mu$, we can reexpress the
hadronic tensor in Eq.~(\ref{w}) as 
\begin{equation}
W^{\mu\nu}(x_B,Q^2) = e^{\mu\nu}_L\, F_L(x_B,Q^2)
                    + e^{\mu\nu}_T\, F_1(x_B,Q^2) \ ,
\label{wlt}
\end{equation}
where the longitudinal and transverse tensors are defined as
\begin{eqnarray}
e^{\mu\nu}_L &\equiv & \frac{1}{2Q^2}\,
\left(x_B\, p\cdot n\, \bar{n}^\mu 
     +\frac{Q^2}{2x_B\, p\cdot n}\, n^\mu \right)
\left(x_B\, p\cdot n\, \bar{n}^\nu 
     +\frac{Q^2}{2x_B\, p\cdot n}\, n^\nu \right) \ ,
\nonumber \\
e^{\mu\nu}_T &\equiv & \bar{n}^\mu\, n^\nu
 + n^\mu\, \bar{n}^\nu - g^{\mu\nu} \equiv d^{\mu\nu}\ ,
\label{elt}
\end{eqnarray}
with $e_{L_{\mu\nu}}e_L^{\mu\nu} = 1/4 $, $e_{T_{\mu\nu}}e_T^{\mu\nu}
= 2$, and $e_{L_{\mu\nu}}e_T^{\mu\nu} = 0$. 
The longitudinal structure function, $F_L(x_B,Q^2)$ in Eq.~(\ref{wlt}), 
is defined as usual, $F_L = F_2/x_B - 2F_1$.  From Eq.~(\ref{wlt}),
the DIS structure functions can be extracted from the hadronic tensor
as  
\begin{eqnarray}
F_1(x_B,Q^2) &=& \frac{1}{2}\, e_T^{\mu\nu}\, W_{\mu\nu}(x_B,Q^2)\,
\nonumber \\
F_L(x_B,Q^2) &=& 4\, e_L^{\mu\nu}\, W_{\mu\nu}(x_B,Q^2)\ .
\label{flt}
\end{eqnarray}

\begin{figure}
\epsfig{figure=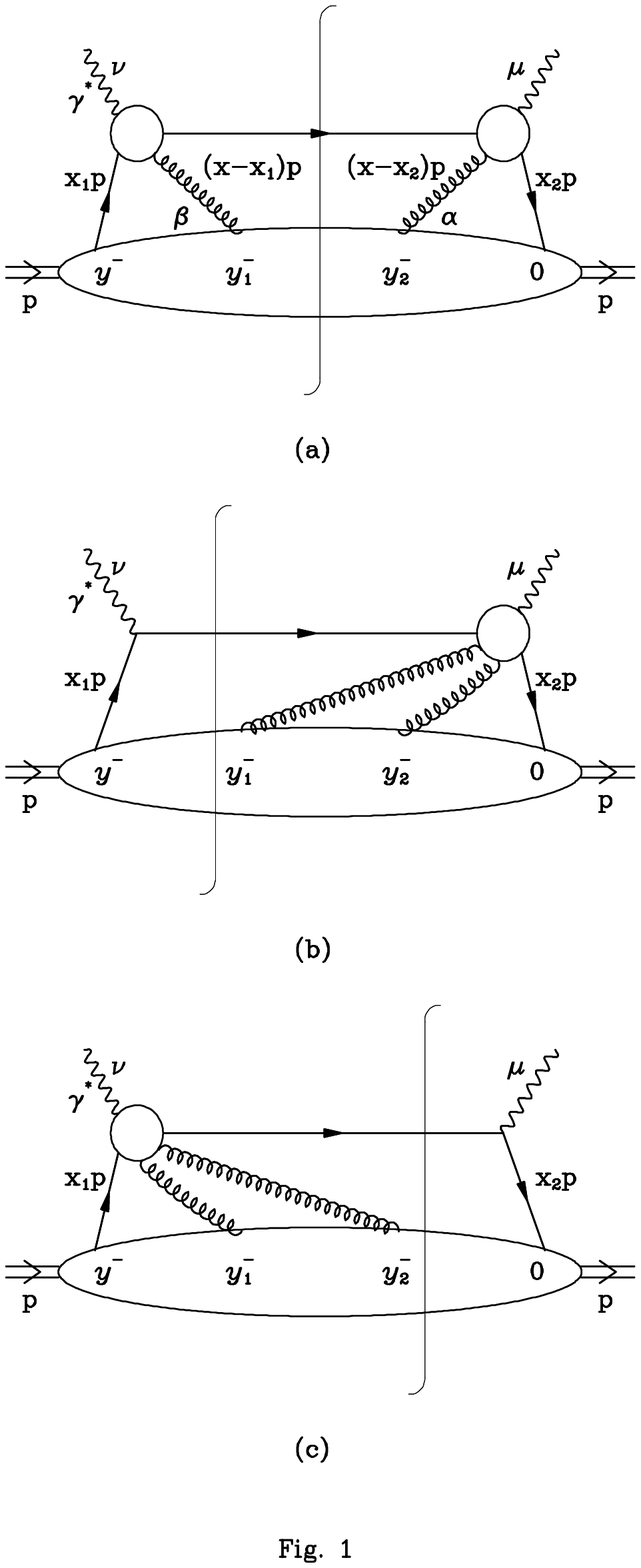,width=2.0in}
\caption{Factorized diagrams for the leading order power corrections
from the quark-gluon subprocesses.}
\label{fig1}
\end{figure}

\begin{figure}
\epsfig{figure=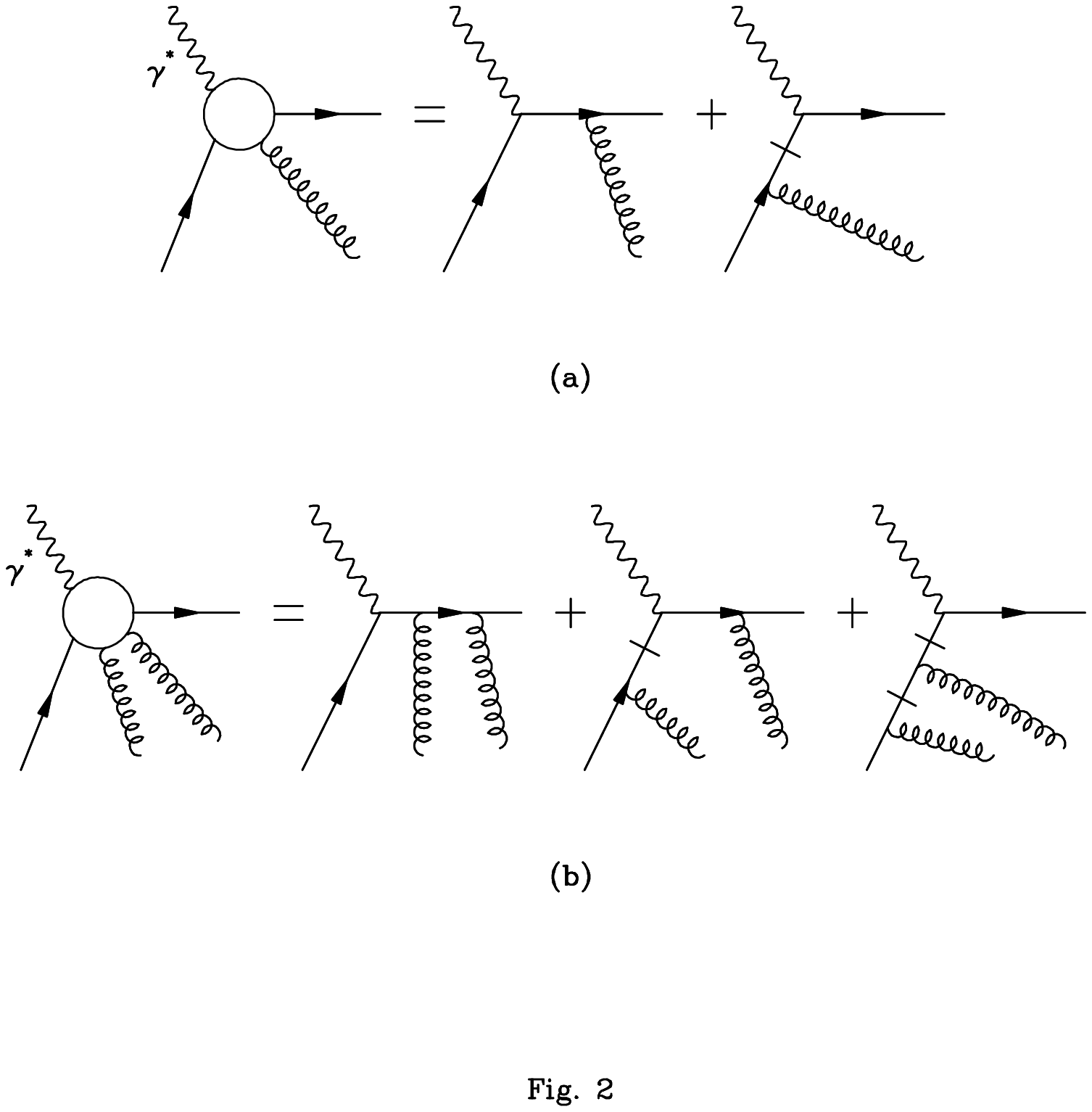,width=2.0in}
\caption{Feynman diagrams for the blobs in Fig.~\protect\ref{fig1}.
The wave lines and curly lines represent the virtual photons and
gluons, respectively.  The solid lines are for the quarks, and the
line with a short bar represents the contact term of the propagator
\protect\cite{Qiu}. }
\label{fig2}
\end{figure}

As derived in Ref.~\cite{Qiu}, the $1/Q^2$ power corrections to the
structure functions from the quark-gluon subprocesses can be
factorized, as shown in Fig.~\ref{fig1},
\begin{equation}
W^{\mu\nu}(x_B,Q^2)\bigg|_{1/Q^2} 
= \frac{1}{Q^2}\, \int dx\, dx_1\, dx_2\,
{\rm Tr}\left[C^{\mu\nu}_{\alpha\beta}(x_B,x,x_1,x_2)\,
T^{\alpha\beta}(x,x_1,x_2,Q^2)\right]\ ,
\label{w4}
\end{equation}
where "Tr" is the spinor trace for fermions.
The partonic part, $C^{\mu\nu}_{\alpha\beta}(x_B,x,x_1,x_2)$ in above
equation, is given by the top parts of the Feynman diagrams in
Fig.~\ref{fig1}.  The blobs in Fig.~\ref{fig1}a are given by the
diagrams in Fig.~\ref{fig2}a, and the blobs in Fig.~\ref{fig1}b and 
Fig.~\ref{fig1}c are given by the diagrams in Fig.~\ref{fig2}b.  In
Fig.~\ref{fig2}, the fermion lines with a short bar indicates that the
propagator includes only the short-distance contact term, as defined
in Ref.~\cite{Qiu}. 
The quark-gluon correlation functions, 
$T^{\alpha\beta}(x,x_1,x_2,Q^2)$ in Eq.~(\ref{w4}), is expressed in
terms of the four-parton matrix element, 
\begin{eqnarray}
T^{\alpha\beta}(x,x_1,x_2,Q^2) &=& \int\frac{p^+dy^-}{2\pi}\,
\frac{p^+dy^-_1}{2\pi}\, \frac{p^+dy^-_2}{2\pi}\,
e^{ix_1p^+y^-}\, e^{i(x-x_1)p^+y^-_1}\, e^{-i(x-x_2)p^+y^-_2}
\nonumber \\
&\times &
\langle p|\bar{\psi}(0)
D^\alpha_T(y_2^-)D^\beta_T(y_1^-)\psi(y^-) |p\rangle\, ,
\label{Tgeneral}
\end{eqnarray}
where $D^\alpha(y) \equiv \partial^\alpha + ig\,T^B A_B^\alpha(y)$ 
is the covariant derivative, which differs by a factor of $i$ from the 
definition used in Refs.~\cite{EFP,Qiu}.  In Eq.~(\ref{Tgeneral}),
$D_T^\alpha \equiv (g^{\alpha\beta} - \bar{n}^\alpha\,
n^\beta)\,D_\beta$, and the line-integrals between the fields are
omitted \cite{Qiu}.

To complete the factorization between the partonic parts and the
twist-4 matrix elements, we need to separate the spinor trace, as well
as the Lorentz indices linking between the $C^{\mu\nu}_{\alpha\beta}$
and $T^{\alpha\beta}$ in Eq.~(\ref{w4}).  Such factorization can be
achieved by decomposing the $T^{\alpha\beta}$ as 
\begin{equation}
T^{\alpha\beta} 
= \frac{1}{2}\,\gamma\cdot p\, d^{\alpha\beta}\, T
+ \frac{1}{2}\,\gamma\cdot p\,\gamma_5\,
  \left(i\epsilon^{\alpha\beta}\right)\, \tilde{T}\, 
+ \dots\ , 
\label{Tab}
\end{equation}
where $T$ and $\tilde{T}$ are scalar functions of $x,x_1,x_2$ and
$Q^2$, and ``\dots'' represents terms that are further power suppressed.
In Eq.~(\ref{Tab}), the symmetric tensor $d^{\alpha\beta}$ is defined
in Eq.~(\ref{elt}) and the antisymmetric tensor
$\epsilon^{\alpha\beta}=\epsilon^{\alpha\beta\mu\nu}\,\bar{n}_\mu 
n_\nu$.  From Eqs.~(\ref{flt}) and (\ref{w4}), we can derive the
$1/Q^2$ contributions to the structure functions at the leading order
in $\alpha_s$ by calculating the Feynman diagrams in Fig.~\ref{fig2}.
Substituting Eq.~(\ref{Tab}) into Eq.~(\ref{w4}), and contracting
$e_{T}^{\mu\nu}$ with the hadronic tensor
$W_{\mu\nu}(x_B,Q^2)|_{1/Q^2}$, we obtain 
in the light-cone gauge ($n\cdot A=0$), 
\begin{mathletters}
\label{etw}
\begin{eqnarray}
e_T^{\mu\nu}\, W_{\mu\nu}^{(a)}(x_B,Q^2)\bigg|_{1/Q^2}
&=& 0 \ ,
\label{etwa} \\
e_T^{\mu\nu}\, W_{\mu\nu}^{(b)}(x_B,Q^2)\bigg|_{1/Q^2}
&=& -e_q^2\ \frac{2x_B}{Q^2}\, \int dx_1\, dx_2\,
    \frac{\delta(x_1-x_B)}{x_2-x_B}\, T_q(x_2,x_1,Q^2)\, ,
\label{etwb} \\
e_T^{\mu\nu}\, W_{\mu\nu}^{(c)}(x_B,Q^2)\bigg|_{1/Q^2}
&=& -e_q^2\ \frac{2x_B}{Q^2}\, \int dx_1\, dx_2\,
    \frac{\delta(x_2-x_B)}{x_1-x_B}\, T_q(x_2,x_1,Q^2)\, ,
\label{etwc}
\end{eqnarray}
\end{mathletters}
where $e_q$ is the quark(or antiquark)'s fractional charge of flavor
$q$, and the superscripts $(a), (b)$ and $(c)$ correspond to the
contributions from diagrams Figs.~\ref{fig2}a, \ref{fig2}b and
\ref{fig2}c, respectively. 
The quark-gluon correlation function $T_q(x_2,x_1,Q^2)$ is defined
as 
\begin{eqnarray}
T_q(x_2,x_1,Q^2) &=& \frac{1}{4}\, \int\frac{dy^-}{2\pi}\,
\frac{p^+ dy^-_1}{2\pi}\,
e^{ix_1p^+y^-}\, e^{i(x_2-x_1)p^+y^-_1}
\nonumber \\
&\times &
\langle p|\bar{\psi}_q(0)\gamma^+\left(d_{\alpha\beta}
D^\alpha(y_1^-)D^\beta(y_1^-)\right) \psi_q(y^-) |p\rangle\, .
\label{T21}
\end{eqnarray}
Substituting Eq.~(\ref{etw}) into Eq.~(\ref{flt}), we obtain
\begin{equation}
F_1(x_B,Q^2)\bigg|_{1/Q^2} = \sum_q e_q^2\ \frac{x_B}{Q^2}\,
\int dx_1\, dx_2\,\left[
    \frac{\delta(x_2-x_B)-\delta(x_1-x_B)}{x_2-x_1}\right] 
    T_q(x_2,x_1,Q^2)\, .
\label{f14t}
\end{equation}
Similarly, by contracting $e_L^{\mu\nu}$ to the hadronic tensor
$W_{\mu\nu}(x_B,Q^2)$, we obtain
\begin{mathletters}
\label{elw}
\begin{eqnarray}
e_L^{\mu\nu}\, W_{\mu\nu}^{(a)}(x_B,Q^2)\bigg|_{1/Q^2}
&=& e_q^2\ \frac{2}{Q^2}\, \int dx\, \delta(x-x_B)\, T_q(x,Q^2)\ ,
\label{elwa} \\
e_L^{\mu\nu}\, W_{\mu\nu}^{(b)}(x_B,Q^2)\bigg|_{1/Q^2}
&=& 0\ , 
\label{elwb} \\
e_L^{\mu\nu}\, W_{\mu\nu}^{(c)}(x_B,Q^2)\bigg|_{1/Q^2}
&=& 0\ ,
\label{elwc}
\end{eqnarray}
\end{mathletters}
where the quark-gluon correlation function $T_q(x,Q^2)$ is defined as 
\begin{equation}
T_q(x,Q^2) = -\frac{1}{4}\, \int\frac{dy^-}{2\pi}\, e^{ixp^+y^-}\, 
\langle p|\bar{\psi}_q(0)\gamma_\alpha D^{\alpha}_T(0)\,
\gamma^+\, \gamma_\beta D^\beta_T(y^-)\psi_q(y^-) |p\rangle\, .
\label{Tx}
\end{equation}
Combining Eqs.~(\ref{flt}) and (\ref{elw}), we have
\begin{equation}
F_L(x_B,Q^2)\bigg|_{1/Q^2} = \sum_q e_q^2\ \frac{8}{Q^2}\,
    T_q(x_B,Q^2)\, .
\label{fl4t}
\end{equation}

From Eqs.~(\ref{f14t}) and (\ref{fl4t}), we obtain the leading order
$1/Q^2$ contributions from quark-gluon subprocess to the structure
function $F_2(x_B,Q^2)$ as 
\begin{eqnarray}
F_2(x_B,Q^2)\bigg|_{1/Q^2}
=\frac{1}{Q^2}\, \sum_q e_q^2\, x_B &&\Bigg(
 2x_B\,\int dx_1\, dx_2\,\left[
       \frac{\delta(x_2-x_B)-\delta(x_1-x_B)}{x_2-x_1}\right] 
       T_q(x_2,x_1,Q^2)
\nonumber \\
&& + 8\, \int dx\ \delta(x-x_B)\, T_q(x,Q^2) \Bigg)\ .
\label{f24t}
\end{eqnarray}
Our results given in Eqs.~(\ref{f14t}), (\ref{fl4t}) and (\ref{f24t})
are consistent with those in Refs.~\cite{EFP,Qiu}, after taking into
account the difference in the definition of the covariant derivative
(a factor of ``$i$'') and an extra ``$-1$'' in the definition of
$T_q(x,Q^2)$ in Eq.~(\ref{Tx}).

\section{Leading Power Corrections at Large $x_B$}
\label{sec3}

In this section, we identify the approximations that enable us to
derive Eq.~(\ref{f24}) from Eq.~(\ref{f24t}), which represents the
complete $1/Q^2$ contributions to $F_2(x_B,Q^2)$ from the leading
order quark-gluon subprocesses. 

From the phenomenological fitting of the dynamical power corrections
\cite{YB2}, $1/Q^2$ contributions to the structure functions are
most important in large $x_B$ region, because of the $1/(1-x_B)$
factor in Eq.~(\ref{hx}).  From the complete leading order $1/Q^2$
contributions in Eq.~(\ref{f24t}), it is nontrivial to conclude that
the $1/(1-x_B)$ behavior exists.  However, under some simple 
approximations, we demonstrate below that Eq.~(\ref{f24t}) has indeed
the $1/(1-x_B)$ behavior when $x_B$ is large.

The second term in Eq.~(\ref{f24t}) is simple and directly
proportional to the quark-gluon correlation function $T_q(x_B,Q^2)$ in
Eq.~(\ref{Tx}).  To see if the $T_q(x_B,Q^2)$ has the $1/(1-x_B)$
behavior, we simplify the operator expression of the $T_q(x_B,Q^2)$ by
using the equation of motion, $\gamma_\alpha D^\alpha(y) \psi_q(y) =
0$, and we have
\begin{eqnarray}
T_q(x_B,Q^2) 
&=& -\frac{1}{4}\, \int\frac{dy^-}{2\pi}\, e^{ix_Bp^+y^-}\, 
\langle p|\bar{\psi}_q(0)\gamma^- D^+(0)\,
\gamma^+\, \gamma^- D^+(y^-)\psi_q(y^-) |p\rangle
\nonumber \\
&=& \frac{(x_B)^2}{2}\, \int\frac{p^+dy^-}{2\pi}\, e^{ix_Bp^+y^-}\, 
\langle p|\bar{\psi}_q(0)\left[p^+\gamma^-\right]\psi_q(y^-) 
|p\rangle
\label{Tx2}
\end{eqnarray}
where the light-cone gauge $n\cdot A=0$ and $D^\alpha_T = 
(g^{\alpha\beta}-\bar{n}^\alpha n^\beta)D_{\beta}$ were used.
On the other hand, the normal twist-2 quark distribution can be
expressed in the light-cone gauge as
\begin{eqnarray}
q(x_B,Q^2) 
&=& \frac{1}{2}\, \int\frac{dy^-}{2\pi}\, e^{ix_Bp^+y^-}\, 
\langle p|\bar{\psi}_q(0)\gamma^+\, \psi_q(y^-) |p\rangle
\nonumber \\
&=& \frac{1}{m_N^2}\, \int\frac{p^+dy^-}{2\pi}\, e^{ix_Bp^+y^-}\, 
\langle p|\bar{\psi}_q(0)\left[p^-\gamma^+\right]\psi_q(y^-)
|p\rangle
\label{qx}
\end{eqnarray}
where $m_N^2 \equiv 2p^+p^-$ is the nucleon mass square.  In order to
relate the correlation function $T_q(x_B,Q^2)$ to the quark
distribution $q(x_B,Q^2)$, we estimate the difference between the
matrix elements $\langle p|\bar{\psi}_q(0)
[p^+\gamma^-]\psi_q(y^-) |p\rangle$ in Eq.~(\ref{Tx2}) and 
$\langle p|\bar{\psi}_q(0)[p^-\gamma^+]\psi_q(y^-)|p\rangle$ in 
Eq.~(\ref{qx}) as follows.  For free field operator $\psi_q$'s on free
light quark states, the difference between these two matrix elements
is the difference between $[p^+\,k^-]$ and $[p^-\,k^+]$ with $k$ the
light quark momentum.  In our frame, we can choose 
\begin{equation}
p^- = m_N^2/(2p^+) \quad \mbox{and} \quad
k^- = m_T^2/(2k^+) \quad \mbox{and} \quad
k^+ = x_B p^+ \ ,
\label{km}
\end{equation}
with the transverse mass of the light quark to be $m_T^2 \equiv
m_q^2+k_T^2$.  If we {\it assume} the free field
relation can be generalized to the full fields, 
we obtain
\begin{eqnarray}
&& \int\frac{p^+dy^-}{2\pi}\, e^{ix_Bp^+y^-}\, 
\langle p|\bar{\psi}_q(0)\left[p^+\gamma^-\right]\psi_q(y^-)
|p\rangle 
\nonumber \\
&& \approx 
\frac{\langle m_T^2\rangle}{m_N^2 (x_B)^2}\, 
\int\frac{p^+dy^-}{2\pi}\, e^{ix_Bp^+y^-}\, 
\langle p|\bar{\psi}_q(0)\left[p^-\gamma^+\right]\psi_q(y^-)
|p\rangle \ .
\label{approx1}
\end{eqnarray}
Substituting Eq.~(\ref{approx1}) into Eqs.~(\ref{Tx2}) and (\ref{qx}),
we derive
\begin{equation}
T_q(x_B,Q^2) \approx \frac{1}{2}\, 
                     \langle m_T^2 \rangle\, q(x_B,Q^2)\ .
\label{Tqq}
\end{equation}
We emphasize here that the importance of Eq.~(\ref{Tqq}) is not
how accurate the approximation is, rather it indicates that the second
term in Eq.~(\ref{f24t}) does not have the $1/(1-x_B)$ enhancement.

\begin{figure}
\epsfig{figure=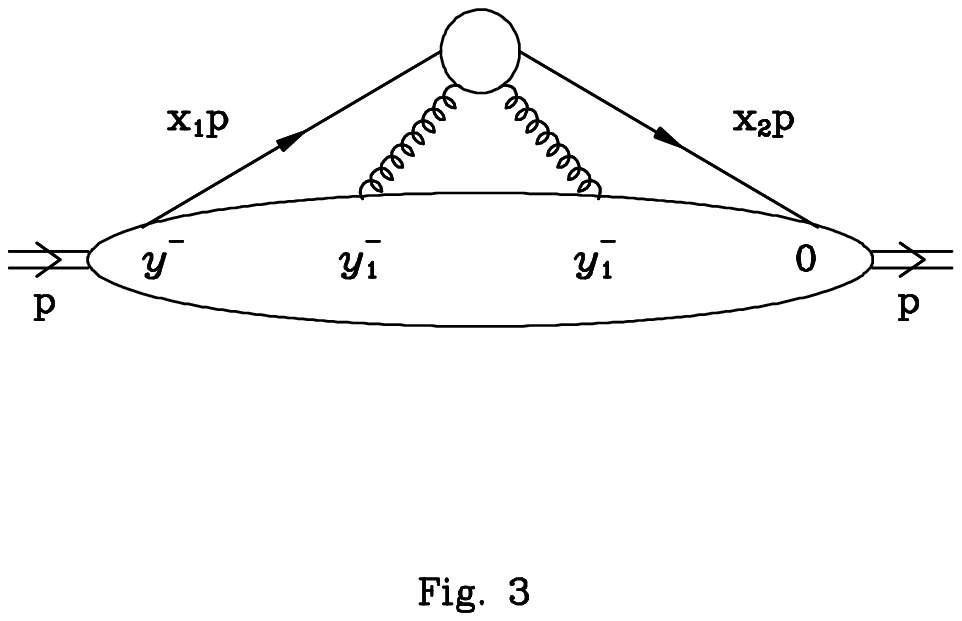,width=2.0in}
\caption{Diagram represents the quark-gluon correlation function
$T_q(x_2,x_1,Q^2)$. }
\label{fig3}
\end{figure}

The first term in Eq.~(\ref{f24t}) depends on the correlation function
$T_q(x_2,x_1,Q^2)$, which can be sketched in Fig.~\ref{fig3}, and be
reexpressed as 
\begin{eqnarray}
T_q(x_2,x_1,Q^2) &=& \frac{1}{2}\, \int\frac{dy^-}{2\pi}\,
e^{ix_1p^+y^-}\, \langle p|\bar{\psi}_q(0)\gamma^+
\nonumber \\
&\times &
\left[ \frac{1}{2}\, \int \frac{p^+ dy^-_1}{2\pi}\,
       e^{i(x_2-x_1)p^+y^-_1}\,
       d_{\alpha\beta}
       D^\alpha(y_1^-)D^\beta(y_1^-)\right] 
\psi_q(y^-) |p\rangle\, .
\label{T21r}
\end{eqnarray}
Because of the factor $[\delta(x_2-x_B)-\delta(x_1-x_B)]/(x_2-x_1)$
in Eq.~(\ref{f24t}), $x_1\sim x_2\sim x_B$ in  $T_q(x_2,x_1,Q^2)$
when $x_B$ is large, and the effective ``gluons'' in Fig.~\ref{fig3} 
are soft.  We then assume that the effective ``gluon'' operator
$d_{\alpha\beta} D^\alpha(y_1^-) D^\beta(y_1^-)$ in Eq.~(\ref{T21r})
is not sensitive to the position variable $y_1^-$, and further
approximate this operator by an averaged expectation value as
\begin{equation}
d_{\alpha\beta} D^\alpha(y_1^-) D^\beta(y_1^-) \approx
\langle d_{\alpha\beta} D^\alpha(0) D^\beta(0) \rangle 
\equiv \langle D_T^2 \rangle\ .
\label{approx2}
\end{equation}
More discussion on this assumption will be given in the 
next section.  With the approximation in Eq.~(\ref{approx2}), we can  
rewrite the correlation function $T_q(x_2,x_1,Q^2)$ as
\begin{eqnarray}
T_q(x_2,x_1,Q^2) &\approx & \frac{1}{2}\, \int\frac{dy^-}{2\pi}\,
e^{ix_1p^+y^-}\, \langle p|\bar{\psi}_q(0)\gamma^+ 
\left[\, \frac{1}{2}\, \langle D_T^2 \rangle\, \delta(x_2-x_1)\right]
\psi_q(y^-) |p\rangle\ ,
\nonumber \\
&=& \left[\, \frac{1}{2}\, \langle D_T^2 \rangle\, \delta(x_2-x_1)
    \right]\, q(x_1,Q^2)\ ,
\label{T21a}
\end{eqnarray}
where $q(x_1,Q^2)$ is the twist-2 quark distribution defined in
Eq.~(\ref{qx}).  Using Eq.~(\ref{T21a}), we derive
\begin{eqnarray}
&& \int dx_1\, dx_2\,\left[
       \frac{\delta(x_2-x_B)-\delta(x_1-x_B)}{x_2-x_1}\right] 
       T_q(x_2,x_1,Q^2)
\nonumber \\ 
&&\quad \approx \frac{1}{2}\, \langle D_T^2 \rangle\, 
  \int_0^1 dx_1\, \delta'(x_1-x_B)\, q(x_1,Q^2)
\nonumber \\
&&\quad  = \frac{1}{2}\,
\langle D_T^2 \rangle \left(-\frac{d}{dx_B}\right) q(x_B,Q^2)\ .
\label{f241}
\end{eqnarray}
Substituting Eqs.~(\ref{Tqq}) and (\ref{f241}) into Eq.~(\ref{f24t}),  
we obtain the leading power corrections to the structure function
$F_2(x_B,Q^2)$ as  
\begin{equation}
F_2(x_B,Q^2)\bigg|_{1/Q^2}
\approx \frac{1}{Q^2}\, \sum_q e_q^2\, x_B \left[
\langle D_T^2 \rangle \left(-x_B\frac{d}{dx_B}\right)
+ 4\, \langle m_T^2\rangle\, \right]\, q(x_B,Q^2)\ .
\label{f24a}
\end{equation}
Combining this leading power corrections and the leading order
twist-2 contributions to $F_2(x_B,Q^2)$, we obtain our main result, 
Eq.~(\ref{f24}). 

Although Eq.~(\ref{f24a}) looks very different from the
phenomenological parameterization in Eq.~(\ref{hx}), we show below
that Eqs.~(\ref{f24a}) and (\ref{hx}) have the same analytical
behavior as $x_B\rightarrow 1$.  When $x_B$ is large, the quark
distribution have the following behavior
\begin{equation}
q(x_B,Q^2) \Longrightarrow (1-x_B)^{n_q(Q^2)}
\quad \mbox{as $x_B\rightarrow 1$} \, , 
\end{equation}
with a positive and real parameter $n_q$.  In general, $n_q(Q^2)$ has
the logarithmic $Q^2$-dependence due to the scaling violation of the
parton distributions.  For the valence quark distributions, $n_q
\sim 3$.  Consequently, we have
\begin{equation}
-x_B \frac{d}{dx_B} q(x_B) =
 n_q(Q^2)\, \frac{x_B}{1-x_B}\, q(x_B,Q^2)\,
+\, \mbox{terms without}\ \frac{1}{1-x_B}\, .
\label{dxqx}
\end{equation}
Substituting Eq.~(\ref{dxqx}) into Eq.~(\ref{f24a}), naturally, we
derive the leading $1/(1-x_B)$ behavior, 
\begin{equation}
F_2(x_B,Q^2)\bigg|_{1/Q^2}
\approx \sum_q e_q^2\, x_B\, q(x_B,Q^2) \left[
\frac{\langle D_T^2 \rangle }{Q^2}\, 
n_q(Q^2)\, \frac{x_B}{1-x_B} 
+\, \mbox{terms without}\ \frac{1}{1-x_B} \right] \, ,
\label{f24x}
\end{equation}
which is consistent with the phenomenological parameterization in
Eq.~(\ref{hx}) that was introduced in Refs.~\cite{YB1,YB2} after
fitting the existing DIS data. 
 
\section{Numerical Results and Conclusions}
\label{sec4}

In this section, we demonstrate numerically that as a function of
$x_B$, our analytical expression in Eq.~(\ref{f24}) and the 
phenomenological parameterization defined in Eqs.~(\ref{f2})
and (\ref{hx}) have almost identical behavior, even though they look
very different.

In the following numerical estimates, we use CTEQ3L parton
distributions of Ref.~\cite{CTEQ3} for the quark distributions in the
proton.  In addition, in order to avoid the resonance region and to
test the perturbative power expansion of the structure functions, we
keep $W^2\geq 2$~GeV$^2$ in our calculations.

\begin{figure}
\begin{minipage}[t]{2.0in}
\epsfig{figure=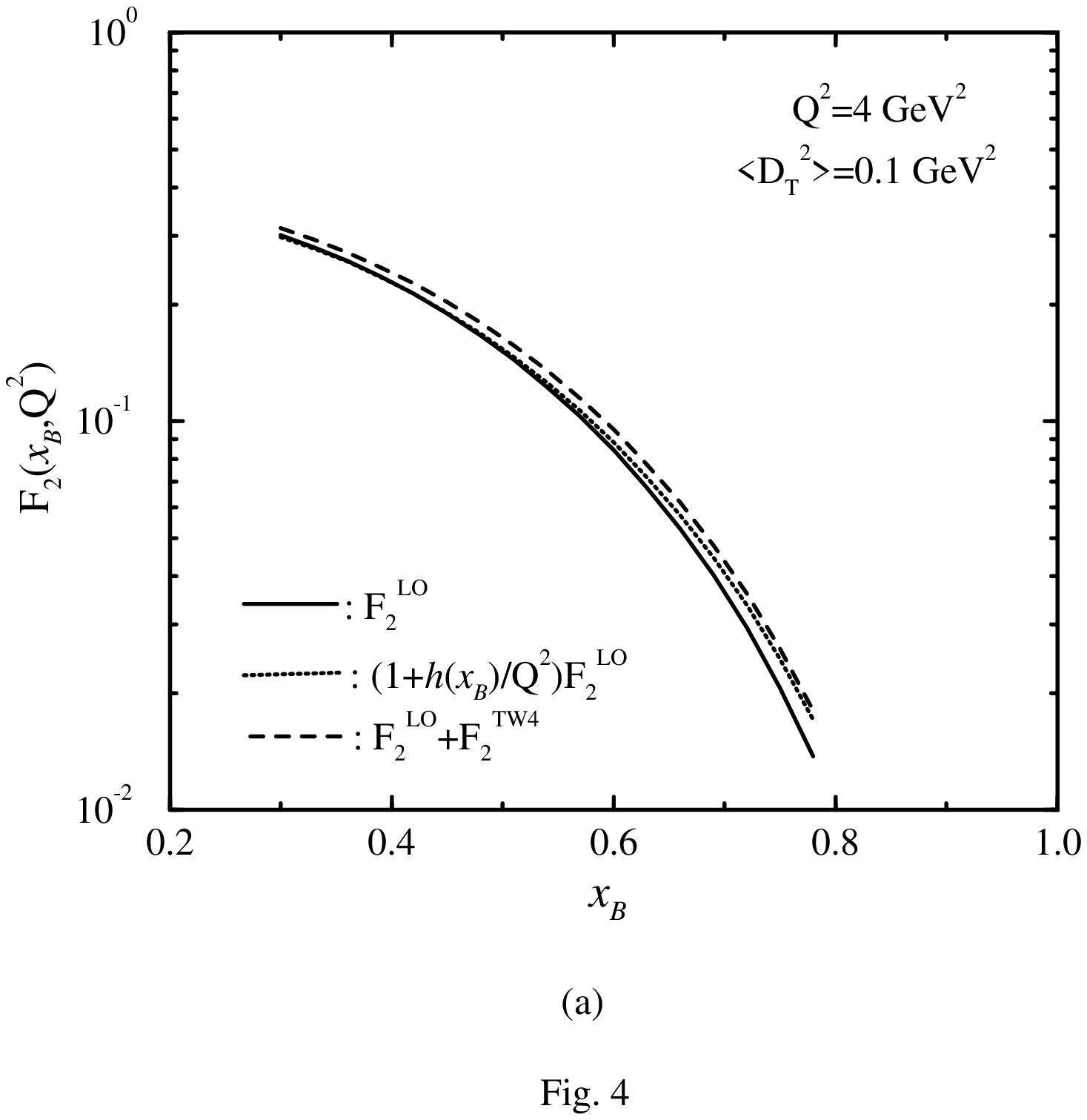,width=1.0in}
\end{minipage}
\hfill
\begin{minipage}[t]{2.0in}
\epsfig{figure=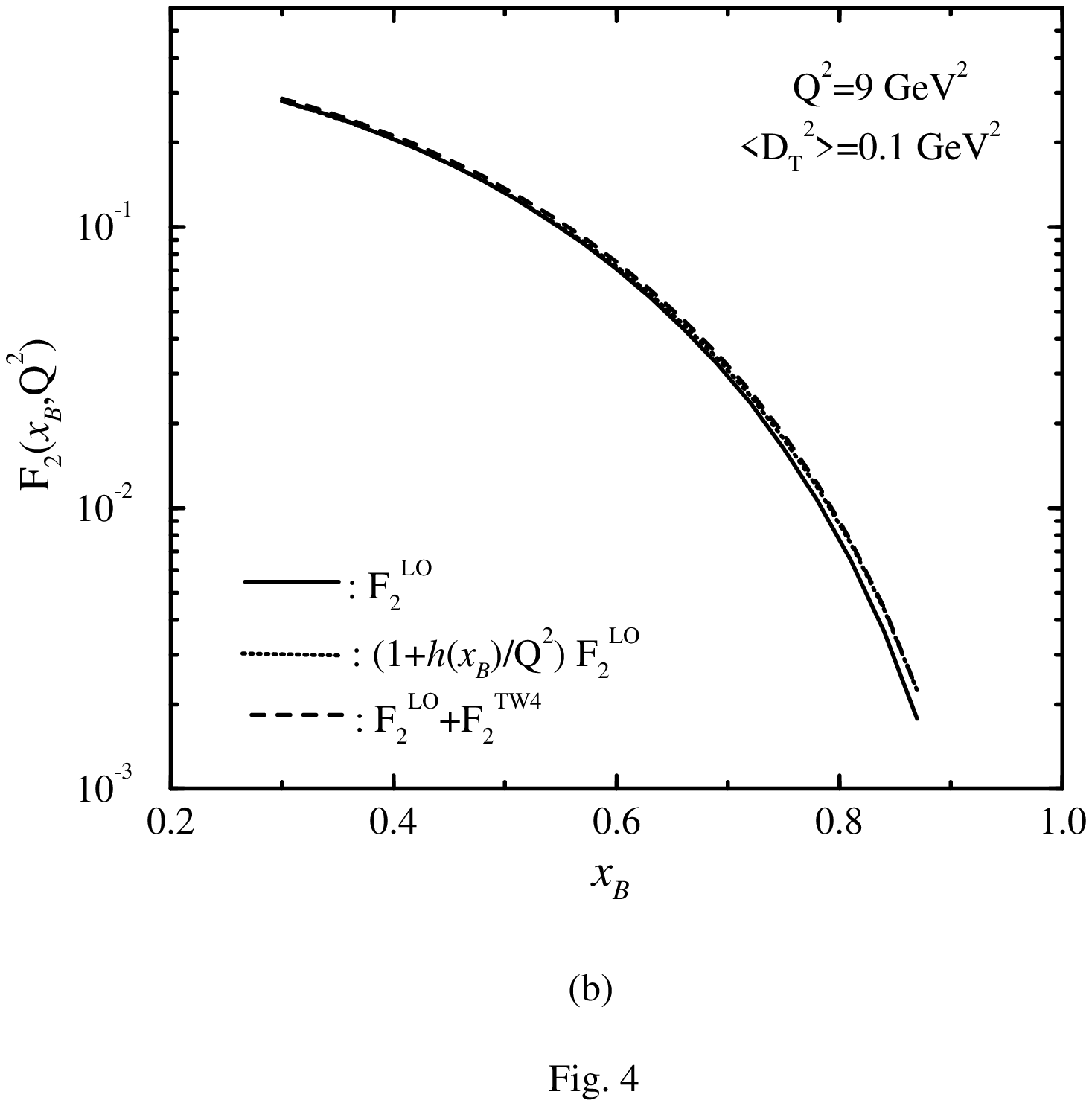,width=1.0in}
\end{minipage}
\caption{Structure function $F_2(x_B,Q^2)$ as a function of $x_B$ at
4~GeV$^2$ (a), and 9~GeV$^2$ (b).}
\label{fig4}
\end{figure}

In Fig.~\ref{fig4}, we plot the structure function $F_2(x_B,Q^2)$ as a
function of $x_B$ for $Q^2=4$ and $9$~GeV$^2$.  The solid line is the
lowest order contribution, $F_2^{\rm LO}(x_B,Q^2)
=\sum_q e_q^2 x_B q(x_B,Q^2)$.
The dotted and dashed lines are given by the expressions in
Eqs.~(\ref{f2}) and (\ref{f24}), respectively, with $F_2^{\rm
LO}(x_B,Q^2)$ for the $F_2(x_B,Q^2)_{\rm LT}$ in Eq.~(\ref{f2}).  
For the fitting parameters, $a,b$ and $c$ in the phenomenological
parameterization $h(x)$, we use the following values $a=0.50$, $b=3.2$
and $c=0.11$, which 
are given in Ref.~\cite{YB2}. For our analytical expression in
Eq.~(\ref{f24}), there are two unknown parameters $\langle D_T^2 
\rangle$ and $\langle m_T^2 \rangle$.  We choose $\langle D_T^2
\rangle =0.1$ GeV$^2$ to be consistent with the overall size of the
power corrections.  The transverse mass square
$\langle m_T^2 \rangle$ for the light quarks should be of the order of
the intrinsic transverse momentum square, and therefore, $\langle 
m_T^2 \rangle \ge \Lambda_{\rm QCD}^2$.  The actual size of $\langle
m_T^2\rangle$ should depend on the collision energy $W^2$ due to
the soft gluon shower from the incoming quark line \cite{CTEQkt}.  
Within our approximation, different choices for the
value of $\langle m_T^2 \rangle$ represents a constant shift of the
size of the leading power corrections in  $(F_2-F_2^{\rm
LO})/F_2^{\rm LO}$.  In the rest of our discussions, we choose
$\langle m_T^2 \rangle$ to be zero for a close exam of the derivative
term. In this paper, we are more interested in understanding the
functional dependence of the power corrections than 
trying to extract the best values of the $\langle D_T^2 
\rangle$ and $\langle m_T^2 \rangle$ by fitting any data.  We defer
the detailed global QCD analysis for extracting the parton
distributions from the data to another publication.

\begin{figure}
\begin{minipage}[t]{2.0in}
\epsfig{figure=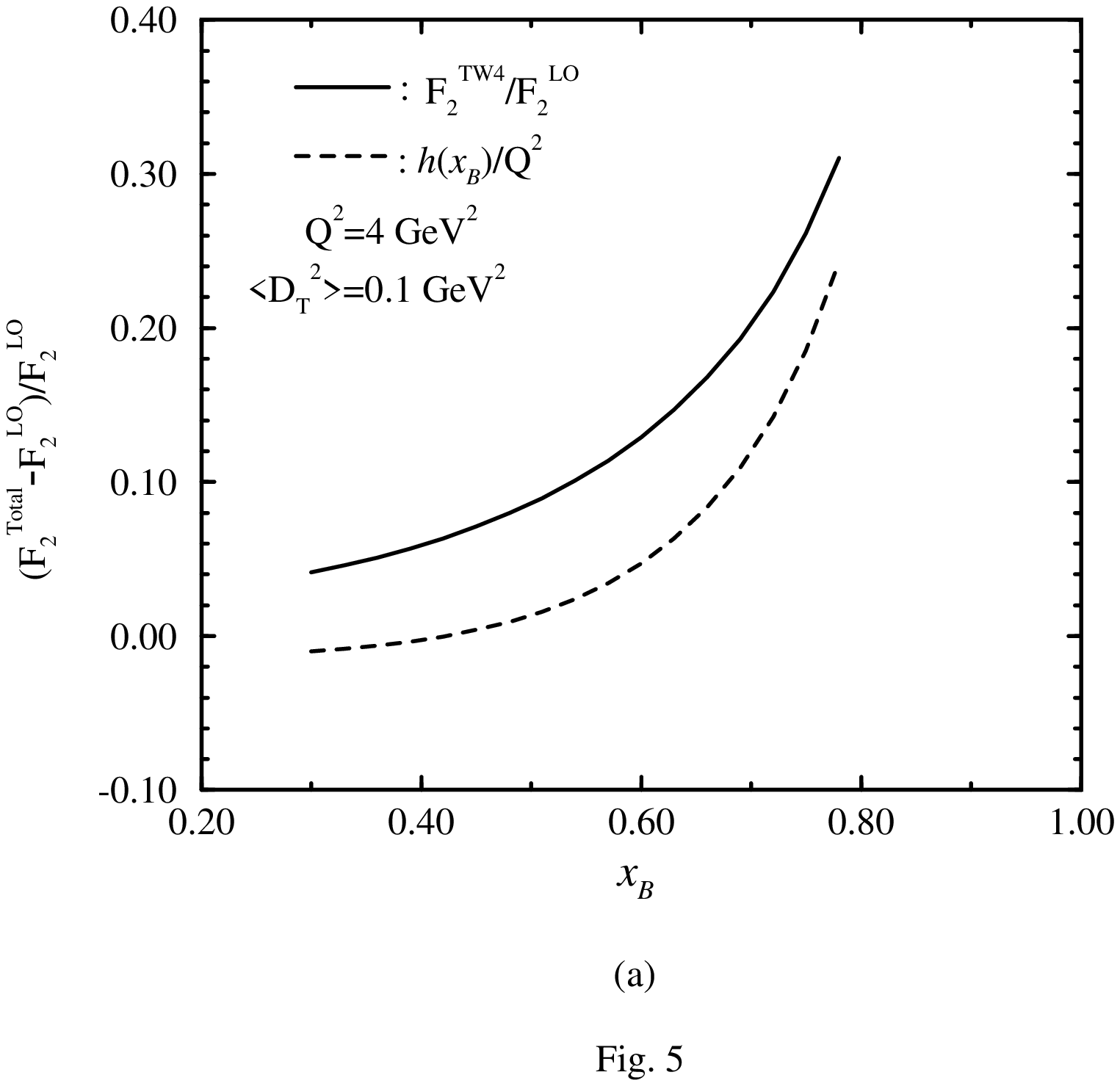,width=1.0in}
\end{minipage}
\hfill
\begin{minipage}[t]{2.0in}
\epsfig{figure=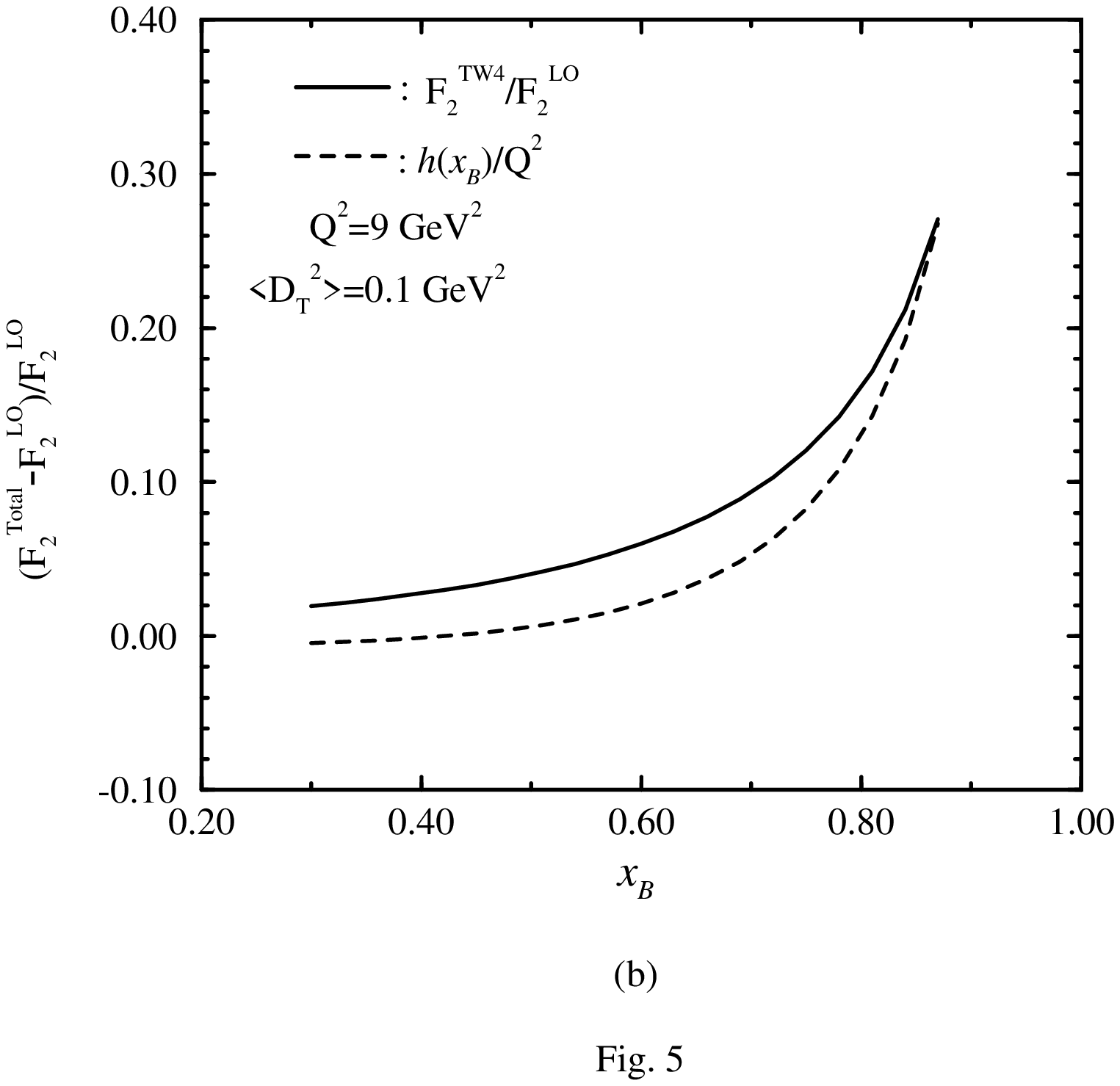,width=1.0in}
\end{minipage}
\caption{Relative power corrections to the structure function
$F_2(x_B,Q^2)$ as a function of $x_B$ at 4~GeV$^2$ (a), and 9~GeV$^2$
(b).} 
\label{fig5}
\end{figure}

Because of the steep falling parton distributions when $x_B$
increases, we plot the relative power corrections to the structure
function, $(F_2-F_2^{\rm LO})/F_2^{\rm LO}$, as a function of $x_B$
in Fig.~\ref{fig5}.  As shown in Fig.~\ref{fig5}, the leading order
power corrections are very significant and can be as large as $30\% $
in large $x_B$ region.  In addition, with only one parameter $\langle
D_T^2 \rangle$ (because $\langle m_T^2 \rangle$ is set to be zero
here),  our analytical result in Eq.~(\ref{f24}) is remarkably
consistent with the parameterization extracted from the data.  The
difference between the solid and the dashed curves in Fig.~\ref{fig5}
cannot be separated by the existing data.  Such consistency on 
$x_B$-dependence is already a nontrivial test for the QCD perturbation
theory. 

\begin{figure}
\epsfig{figure=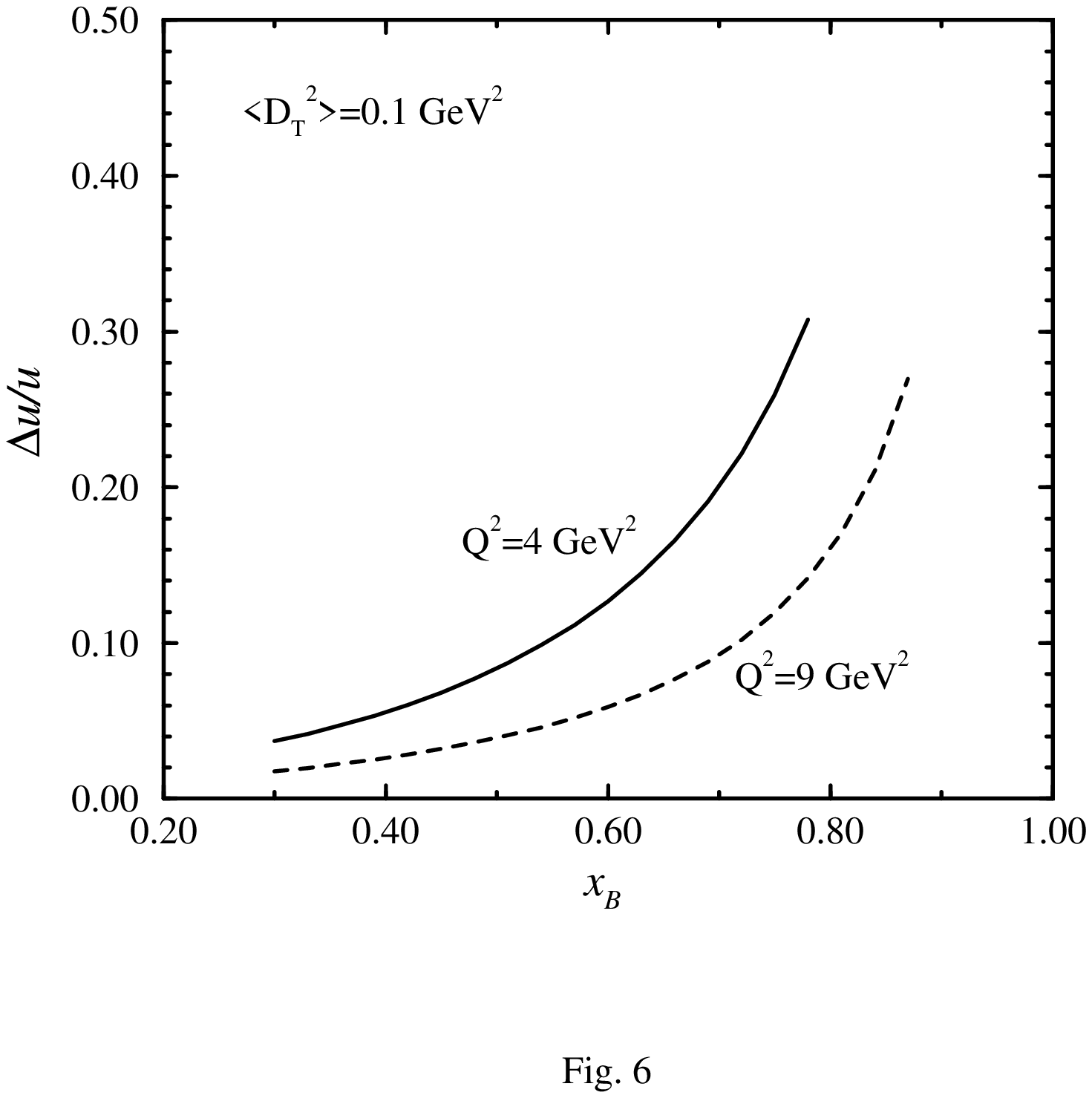,width=1.0in}
\caption{Relative power corrections to the ``up'' quark distribution
at $Q^2=4$~GeV$^2$ and $Q^2=9$~GeV$^2$, respectively. }
\label{fig6}
\end{figure}

\begin{figure}
\epsfig{figure=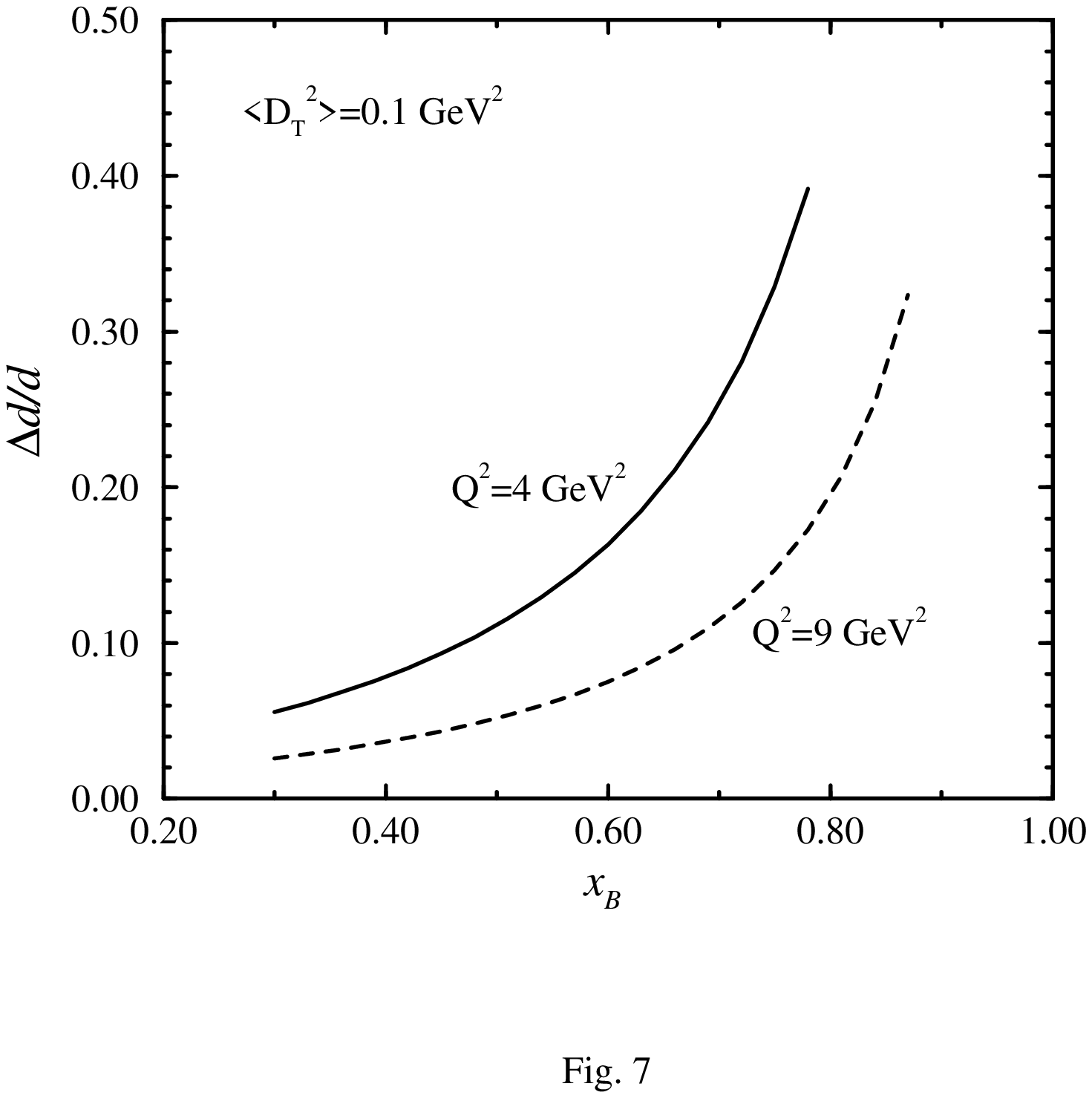,width=1.0in}
\caption{Relative power corrections to the ``down'' quark distribution
at $Q^2=4$~GeV$^2$ and $Q^2=9$~GeV$^2$, respectively. }
\label{fig7}
\end{figure}

As shown in last section, the leading power corrections at
large $x_B$ are given by the derivatives of the normal quark
distributions.  It is the derivative that makes the power corrections
different for quark distributions of different flavors.  We introduce
\begin{equation}
\Delta q(x_B,Q^2) \equiv \frac{1}{Q^2}\, \left[
\langle D_T^2 \rangle \left(-x_B\frac{d}{dx_B}\right)
+ 4\langle m_T^2\rangle \right]\, q(x_B,Q^2)\ ,
\label{dq}
\end{equation}
for quark flavor $q$.  In
Figs.~\ref{fig6} and \ref{fig7}, we plot the relative power
corrections $\Delta q/q$ as a function of $x_B$ for the ``up'' and 
``down'' quark, respectively.  
Because the ``down'' quark
distribution falls off faster than the ``up'' quark distribution in
the CTEQ3L parton distributions, the ``down'' quark obtains a larger
power correction than what the ``up'' quark gets.  Such difference 
is clearly evident in Figs.~\ref{fig6} and \ref{fig7}, and is very
significant when $x_B$ is large.  
The power corrections derived here also receive some logarithmic
$Q^2$-dependence through the derivative of the parton distributions,
due to the scaling violation of the twist-2 parton distributions.  

As demonstrated above, the power corrections to the structure
functions are not only significant at large $x_B$, but also 
sensitive to the quark flavors.  It is such flavor difference of
the power corrections that makes the extraction and the flavor
separation of the parton distributions nontrivial, in particular, in
large $x_B$ region.  Recently, a lot of attention have been devoted to
the ratio of $d/u$ \cite{YB1}.  We defer the detailed global
QCD analysis for the extraction and the flavor separation of the
parton distributions to another publication. 

In summary, we derive the analytical expressions for the leading power
corrections to the DIS structure function, given in
Eq.~(\ref{f24a}).  We made two main 
assumptions in our derivation, which are introduced in
Eqs.~(\ref{approx1}) and (\ref{approx2}), respectively. 
Since the $1/(1-x_B)$ enhancement of
the power corrections is from the term proportional to the quark-gluon
correlation $T_q(x_2,x_1,Q^2)$, the assumption in Eq.~(\ref{approx2})
is more crucial than the other.  In general, it may be possible that
the $\langle D_T^2 \rangle$ is not a constant independent of the
position (or equivalently the value of $x_B$).  However, as long as
any part of the $\langle D_T^2 \rangle$ is not proportional to
$(1-x_B)$, we will have the $1/(1-x_B)$ enhancement.  We believe that
at large $x_B$, at least the ``soft'' gluon part of the averaged value
of the covariant derivative 
square should not be sensitive the location inside the nucleon, even
though the averaged value of the transverse momentum square may be
proportional to $(1-x_B)$.  Therefore, we conclude that the leading order
power corrections to the structure function $F_2(x_B,Q^2)$ should have
the $1/(1-x_B)$ enhancement. and our analytical result,
Eq.~(\ref{f24}) with two fitting parameters $\langle D_T^2 \rangle$
and $\langle m_T^2 \rangle$,
should predicts the correct {\it leading} $x_B$-dependence of the
power corrections at large $x_B$.  Because it has only two 
unknown parameters, our result has a good predicting power and can 
be tested at CEBAF as well as future fixed target programs at
Fermilab.  It is also useful for the QCD global analysis for
extracting parton distributions in large $x_B$ region.  In addition, 
our approach can be applied to the Drell-Yan process to study the
leading power corrections there \cite{QS}.


\section*{Acknowledgment}

We are pleased to acknowledge useful discussions with C. Keppel
and W. Zhu.  This work was supported in part by the U.S. Department of
Energy under Grant Nos. DE-FG02-87ER40731 and DE-FG02-96ER40989, and
under the Contract No. DE-AC02-98CH10886.



\end{document}